\documentclass[fleqn,twoside]{article}
\usepackage{espcrc2}

\usepackage{graphicx}
\usepackage[figuresright]{rotating}

\def\lapproxeq{\lower .7ex\hbox{$\;\stackrel{\textstyle
<}{\sim}\;$}}
\def\gapproxeq{\lower .7ex\hbox{$\;\stackrel{\textstyle
>}{\sim}\;$}}

\newcommand{\AmS}{{\protect\the\textfont2
  A\kern-.1667em\lower.5ex\hbox{M}\kern-.125emS}}

\title{Spin structure function $g_1$ at low $x$: status and plans}

\author{B. Bade\l ek\address[MCSD]{Physics Department, Uppsala University, Box 530,
        S - 751 21 Uppsala, Sweden, and \\
        \mbox{}\hspace{2mm}Institute of Experimental Physics, 
        Warsaw  University, Hoza 69, PL - 00 681 Warsaw, Poland}%
        \thanks{Supported in part by the KBN grants 2 P03B 05119 and 
\mbox{}\hspace{1.5mm}SPUB nr 621/E-78/SPUB-M/CERN/P-03/DZ298/2000}}
       
\begin{document}

\begin{abstract}
A brief review of measurements and expectations concerning the spin structure function
$g_1$ of the nucleon at low values of the scaling variable $x$ is given.
\vspace{1pc}
\end{abstract}

\maketitle

\section{INTRODUCTION}
The spin-dependent structure function $g_1$ is specially
interesting at low $x$, i.e. at high parton densities, where new dynamical
mechanisms may be revealed. Its knowledge there is needed for
evaluating the spin sum rules ne\-ces\-sary to understand
the origin of nucleon spin. 
The behaviour of $g_1$ at $x\lapproxeq$ 0.001 and in the scaling region, $Q^2
\gapproxeq$1 GeV$^2$, 
is unknown. Spin independent structure function $F_2$ rises there,
in agreement with QCD and contrary to the Regge model predictions. 
A great opportunity to explore the spin dependent phenomena in the 
region of low $x$ would be through polarising the proton beam at HERA.
In the fixed target experiments, low values of $x$ are correlated
with low values of $Q^2$.
Theoretical analysis of these results thus requires a suitable
extrapolation of $g_1$ to the low $Q^2$ region
where the dynamical mechanisms, like the Vector Meson Dominance (VMD),
 can be important.
For large $Q^2$ the VMD contribution to $g_1$ vanishes as $1/Q^4$
 and can usually be neglected.
Moreover, the partonic contribution to $g_1$
 which controls the structure functions in the deep inelastic domain
 and which scales there {\it modulo} logarithmic corrections, has to be
 suitably extended to the low $Q^2$ region. 
In the $Q^2$=0 limit $g_1$ should be a finite function of $W^2$,
free from any kinematical singularities or zeros.

\section{PRESENT INFORMATION ON THE $g_1$ OF THE NUCLEON}
\subsection{Results of measurements}
As a result of a large experimental effort over the years, proton and
deuteron $g_1$ was measured for 0.000 06 $< x <$ 0.8, cf. Fig. 
1, \cite{bbjkjk} (and refe\-ren\-ces therein).
Direct measurements on the neutron are limited to $x\gapproxeq$ 0.02 (see 
e.g. \cite{hermes}).
Recent measurements by HERMES \cite{hermes}
are not shown in Fig. 1; those results
cover the region  $x\gapproxeq$ 0.005 and do not change the overall picture.
The region of lowest $x$ values was
explored only by the SMC due to the high energy 
of the muon beam, the demand of a final state hadron in the analysis \cite{smchad} 
and the implementation of a dedicated low $x$ trigger with a calorimeter signal
 \cite{smct15}, the two latter requirements efficiently removing the dominant 
$\mu e$ scattering background. No significant spin effects were observed there.

\begin{figure}[htb]
\includegraphics*[width=6cm]{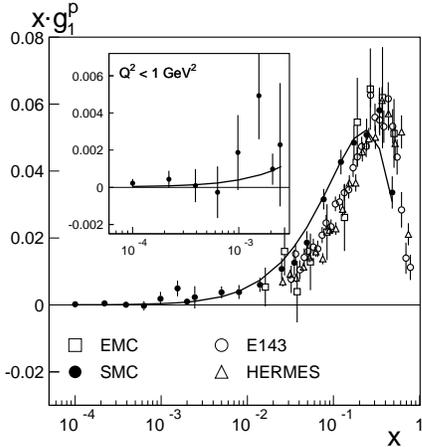}
\vspace{-5pt}
\caption{Summary of the $xg_1^{p}$ as a function of $x$ at the
{\it measured} $Q^2$.
The insets show the SMC data for $Q^2 <$ 1 GeV$^2$. Errors are statistical.
The curves, calculated at ($x,Q^2$) values of the SMC data, result of the model
described in Section 2.5. Figure taken from \cite{bbjkjk}.}
\label{fig:xg1p}
\end{figure}
%

%

As seen in Fig.1 the scaling violation in $g_1(x,Q^2)$  
is weak: the average $Q^2$ is about 10 GeV$^2$ for the SMC
and almost an order of magnitude less for the SLAC and HERMES experiments.
For the SMC data \cite{smct15}, $\langle x \rangle$ = 0.0001 corres\-ponds to 
$\langle Q^2 \rangle$ = 0.02 GeV$^2$; $Q^2$ becomes larger than 1 GeV$^2$
for $x\gapproxeq$ 0.003. 
At low $x$ results on $g_1$ have large errors but it seems that 
$g_1^p$ is positive 
and $g_1^d$ and $g_1^n$ are negative there. Statistical errors dominate 
in that kinematic interval.
It should be noted that a direct result of all measurements is the 
longitudinal cross section asymmetry, $A_{\|}$ which permits to
extract the virtual photon -- proton asymmetry, $A_1$ and finally,
using $F_2$ and $R$, to get $g_1$.

\subsection{Regge pole model expectations}

The low $x$ behaviour of $g_1$ for fixed $Q^2$ reflects the high energy behaviour
of the virtual Compton scattering cross section with centre-of-mass 
energy squared, $W^2$. This is the Regge limit of the (deep) inelastic scattering
where the Regge pole exchange model  should be applicable. 
According to this model, $g_1(x,Q^2) \sim x^{-\alpha}$ for
$x \rightarrow 0$ and fixed $Q^2$, where
$\alpha$  is the intercept of the Regge trajectory, here corresponding
to axial vector mesons.  It is expected that $\alpha \sim 0$ for both
$I=0$ and $I=1$ trajectories, \cite{hei}.
This behaviour of $g_1$ should go smoothly to the $W^{2 \alpha}$
dependence for $Q^2 \rightarrow 0$. It is also expected that
the flavour singlet part of the $g_1$ should have a similar low $x$ behaviour
as the nonsinglet one.
Regge model predictions for $g_1$ become unstable against the QCD evolution which
generates more singular behaviour of $g_1$ than that given by  
$x^{-\alpha}$.

Other predictions based on the Regge model are:
 $g_1\sim$ {ln}$x$, \cite{clo_rob} and
$g_1\sim$ 2~{ln}(1/$x$)--1, \cite{bass_land}. A perverse behaviour,
$g_1\sim$1/$(x$ln$^2x)$, recalled in \cite{clo_rob}, is not valid
for $g_1$, \cite{misha}.

The Regge model prediction, $g_1\sim x^0$, has in the past often been
used to obtain the $x\rightarrow$ 0 extrapolation of $g_1$ (see Fig. 2) 
required to extract its first moments (cf. \cite{qcd_old} and references therein).

\begin{figure}[htb]
\includegraphics*[width=6cm]{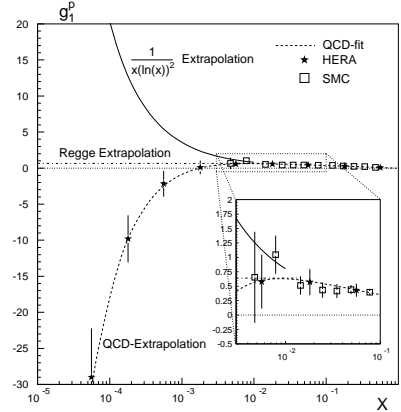}

\vspace{-9pt}
\caption{Three scenarios of the possible behaviour of $g_1^p$ at low $x$
\cite{hera_study}.}
\label{fig:g1extrapo}
\end{figure}

\vspace{-5pt}
The low $x$ data of the SMC \cite{smct15} include the kinematic region where
$W^2=(p+q)^2$ is high, $W^2$\gapproxeq 100 GeV$^2$, and $W^2\ll Q^2$.
Thus the Regge model should be applicable there.
However for those data $W^2$ changes very little:
from about 100 GeV$^2$ at $x=$ 0.1 to about 220 GeV$^2$ at $x=$ 0.0001,
contrary to a strong change of $Q^2$: from about 20 GeV$^2$
to about 0.01 GeV$^2$ respectively.
Thus those data cannot test the Regge
behaviour of $g_1$ through the $x$ dependence of the latter. 

\subsection{QCD fit to the world data on $g_1^N$} 

Next-to-leading (NLO) order QCD analyses of the $Q^2$ dependence of $g_1$
had been performed on the $g_1$ data \cite{qcd_old,roland,qcd_bb} and have 
indicated a large contribution of gluon polarisation to the proton spin 
albeit determined with very large errors.

Extrapolations of QCD fit results to the unmeasured low $x$ region give
 all three structure functions, $g_1^p$, $g_1^d$ and $g_1^n$ 
 negative ($g_1^p$ becomes such below the
lowest $x$ data point used in the fit), due to a large negative singlet
contribution. It should be stressed that 
the $g_1$ results  for $x$ values below these of the data do not influence
the results of the fit. Thus there is no reason to expect that
the partons at very low $x$ behave as those in the measured (larger $x$) region.
Nevertheless the low $x$  QCD extrapolations are 
presently being used to get the $x\rightarrow$ 0 extrapolation of $g_1$
 \cite{qcd_old}, necessary to evaluate its first moments. They  
strongly disagree with the Regge asymptotic form, cf. Fig. 2.

\vspace{-5pt}

\subsection{ln$^2(1/x)$ corrections to $g_1$}

The small $x$ behaviour of $g_1(x,Q^2)$ is controlled by the double
logarithmic ln$^2(1/x)$ contributions, i.e. by those terms of the perturbative
expansion which correspond to the powers of ln$^2(1/x)$ at each order 
of the expansion \cite{bartels}.
The  ln$^2(1/x)$ effects go beyond the standard (i.e. ln$Q^2$) 
LO (and NLO) QCD evolution.
It is convenient to resum these terms 
using the unintegrated (spin-dependent) parton
distributions. 
The resulting integrated parton distributions contain nonperturbative
parts, $\Delta p^0_j(x)$,
 corresponding to low transverse momenta of partons,
to be parametrised semiphenomenologically.

A complete formalism incorporating the LO Altarelli--Parisi evolution
and the  ln$^2(1/x)$ resummation at low $x$ has been built for $g_1^p$ 
\cite{jkbb_p,jkbz_p}. It has been shown that the nonsinglet part of $g_1$
is dominated by ladder diagrams while a contribution of the nonladder
bremsstrahlung diagrams is important for the singlet part. Results are 
presented in Fig. 3; small $x$ effects suppress the very strong
$x$ dependence of $g_1$ resulting from pure Altarelli--Parisi evolution.

\begin{figure}[htb]
\includegraphics*[width=7cm]{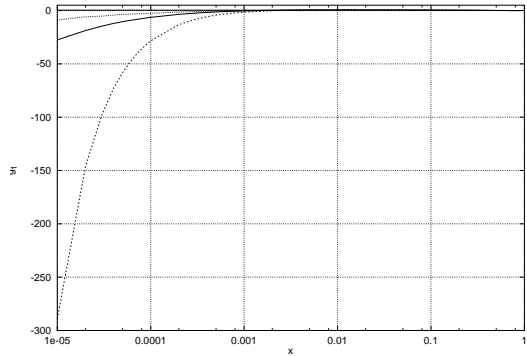}

\vspace{-5pt}
\caption{$g_1^p(x,Q^2)$ at $Q^2$=10 GeV$^2$ after including
the ln$^2(1/x)$ corrections. A thick solid line 
corresponds to full calculations, a dashed one -- only the ladder ln$^2(1/x)$
resummation with LO Altarelli-Parisi evolution, a dotted one - pure LO
Altarelli-Parisi evolution and a thin solid line - the nonperturbative
input to $g_1$ \cite{jkbz_p}.}

\label{jkbz}
\end{figure}

Even if the ln$^2(1/x)$ effects are not important
in the $x$ range of the fixed target experiments
they significantly affect $g_1$ in the low $x$ region
which may be probed at the polarised HERA.

\subsection{Nonperturbative effects in $g_1$}
 
At the low $x$, low $Q^2$ region, $g_1$ was represented by \cite{bbjkjk}:
\begin{equation}
g_1(x,Q^2)=g_{1}^{VMD}(x,Q^2) + g_{1}^{part}(x,Q^2)
\label{g1tot}
\end{equation}
The partonic contribution, $g_1^{part}$ is at low $x$
controlled by the ln$^2(1/x)$ terms resummed using unintegrated
parton distributions, cf. section 2.4. 
This formalism is very suitable for extrapolating
$g_1$ to the region of low $Q^2$ at fixed $W^2$.
Results for $ g_{1}^{part}(x,Q^2)$ are shown as a curve in Fig.1.
The VMD contribution, $g_1^{VMD}(x,Q^2)$, was taken as:
\begin{equation}
g_{1}^{VMD}(x,Q^2)  = {pq\over 4 \pi} \sum_{v=\rho,\omega,\phi}
{ m_v^4 \Delta \sigma_{v}(W^2) \over \gamma_v^2
(Q^2+m_v^2)^2}
\end{equation}
\vskip-0.1cm
where
$\gamma_v^2$ are determined from the leptonic widths of the vector mesons
and $m_v$ is the mass of the vector meson $v$.
Unknown vector meson -- nucleon cross sections
 $\Delta \sigma _{v}=(\sigma_{1/2} - \sigma_{3/2})/2$ (subscripts
refer to the projections of the total spin on the vector meson momentum)
were taken proportional (with a proportionality coeffcient $C$)
to the appropriate combinations of the nonperturbative contributions
$\Delta p_j^0(x)$ to the polarised quark and antiquark distributions.
The $\Delta p^0_j(x)$ behave as $x^0$ for $x\rightarrow$0.
As a result the  cross sections $\Delta \sigma_{v}$ behave as
$1/W^2$ at large $W^2$ that corresponds to zero intercepts of
the appropriate Regge trajectories.
Exact $x$ dependence of $\Delta p_j^0(x)$ was included.

The statistical accuracy of the SMC data is too poor to constraint
the value of the coefficient $C$.
The SLAC E143 data \cite{e143} preferred a small
negative value of C which is consistent with the results of
the phenomenological analysis of the sum rules \cite{IOFFE}.
Similar analysis of the neutron and deuteron spin
structure functions
was inconclusive.

\section{$g_1$ AT FUTURE COLLIDERS}

\subsection{$g_1^p$}

Presently the form of the $x\rightarrow$ 0 extrapolation of the $g_1$
provides the largest error on its first moment: the contribution
from the unmeasured low $x$ region amounts from about 2 to 10$\%$ of the
integral in the measured range, depending whether the Regge model
or the QCD is used for the extrapolation (cf. \cite{qcd_old} and 
references therein). An empirical insight into the validity of different 
low $x$ mechanisms would be possible through polarising 
the proton beam at HERA. This would extend the measured region
of $x$ down to approximately 0.000 06 at $Q^2\gapproxeq$ 1 GeV$^2$.

Polarised $ep$ ($eA$) colliders (HERA, EIC) will also have a potential to explore
the transition between the photoproduction and scaling region,
0$< Q^2 \lapproxeq$  1 GeV$^2$ \cite{hera_study_lowq,eic}. 
Photoproduction measurements would
constraint the spin dependent Regge model and Regge contribution to the
Drell-Hearn-Gerasimov sum rule. Measurements in the transition region would
impose new limits on $g_1$ models containing both perturbative and nonperturbative 
components.

\subsection{$g_1^\gamma$}

Spin dependent structure function of the photon, $g_1^\gamma$,
 can in principle become accessible 
in the future linear $e^+e^-$ or $e\gamma$ linear colliders, \cite{strat}.
The latter mode would be particularly suitable for probing the photon 
structure at low values of $x$. The $g_1^\gamma (x,Q^2)$ has been analysed
\cite{jkbz_g} within the formalism 
being an extension of that developed
for the $g_1^p$i, \cite{jkbz_p}. It was found that e.g. different scenarios 
for the 
nonperturbative spin dependent gluon content of the photon 
give significantly different values of $g_1^\gamma$ at $x\sim$10$^{-5}$.

\section{OUTLOOK}
The spin dependent structure function $g_1$ at low $x$ is both fascinating in
itself and necessary for understanding the origin of the proton spin.
Future colliders, HERA and EIC (and  $e\gamma$ one for the {$g_1^\gamma$})
would give us a great opportunity to explore it.

\end{document}